\documentclass[twocolumn]{aastex631}

\usepackage{multirow}
\usepackage{booktabs}

\begin{document}

\title{Minor-merger induced star formation rejuvenation in an elliptical radio-loud quasar host, 3C 59}

\correspondingauthor{Tao Wang}
\email{taowang@nju.edu.cn}

\author[0000-0002-1010-7763]{Yijun Wang}
\affiliation{School of Astronomy and Space Science, Nanjing University, 163 Xianlin Avenue, Nanjing 210023, People's Republic of China}
\affiliation{Key Laboratory of Modern Astronomy and Astrophysics, Nanjing University, Ministry of Education, 163 Xianlin Avenue, Nanjing 210023, People's Republic of China}

\author[0000-0002-2504-2421]{Tao Wang}
\affiliation{School of Astronomy and Space Science, Nanjing University, 163 Xianlin Avenue, Nanjing 210023, People's Republic of China}
\affiliation{Key Laboratory of Modern Astronomy and Astrophysics, Nanjing University, Ministry of Education, 163 Xianlin Avenue, Nanjing 210023, People's Republic of China}
\thanks{\email{taowang@nju.edu.cn}}

\author{Ke Xu}
\affiliation{School of Astronomy and Space Science, Nanjing University, 163 Xianlin Avenue, Nanjing 210023, People's Republic of China}
\affiliation{Key Laboratory of Modern Astronomy and Astrophysics, Nanjing University, Ministry of Education, 163 Xianlin Avenue, Nanjing 210023, People's Republic of China}

\author{Junjie Mao}
\affiliation{Department of Astronomy, Tsinghua University, Haidian DS 100084 Beijing, People's Republic of China}

\author{Yerong Xu}
\affiliation{Department of Astronomy \& Physics, Saint Mary's University, 923 Robie Street, Halifax, NS B3H 3C3, Canada}
\affiliation{Istituto Nazionale di Astrofisic -- Istituto di Astrofisica Spaziale e Fisica Cosmica di Palermo, Via U. La Malfa 153, I-90146 Palermo, Italy}

\author{Zheng Zhou}
\affiliation{Department of Astronomy, Xiamen University, Xiamen, Fujian 361005, People's Republic of China}




\begin{abstract}

We report a rare case where an elliptical radio-loud quasar host, 3C 59, rejuvenates star formation activity through minor mergers with its nearby satellite galaxies. The inferred star formation history of 3C 59 shows significant star formation rejuvenation within the past 500 Myr, before which remains rather quiescent for most of the cosmic time. Three nearest satellite galaxies of 3C 59 exhibit significant morphological disturbances, and two of them present strong tidal tails pointing towards 3C 59. In addition, all the satellite galaxies within a projected distance of 200 kpc show low star formation activities. They also have systematically lower effective radius ($R_{\rm e}$) than local late-type galaxies, while 3C 59 has significantly larger $R_{\rm e}$ than both early- and late-type galaxies. All these features suggest that ongoing minor mergers between 3C 59 and its nearby satellites could be causing gas to flow into 3C 59, which  induces the star formation rejuvenation, and possibly also triggers the quasar activity. The enormous power from the large-scale radio jet of 3C 59 may in turn help keep the halo hot,  prevent gas cooling, and further reduce star formation in its satellite galaxies. These results provide important insights into the mass and size growth of central galaxies and star formation quenching of satellite galaxies in galaxy groups.

\end{abstract}

%% find keywords in this website: https://uat.astrothesaurus.org/uat/456?view=hierarchy&path=563-573


% #####################################################################
% ########################      Introduction     ##############################
% #####################################################################
\section{Introduction} \label{sec:intro}

Galaxies are mainly divided into two different populations in the star-formation rate vs. stellar mass diagram:
star-forming galaxies (SFGs) with evident star formation and quiescent galaxies (QGs) with little or no star formation
\citep[e.g.,][]{Noeske2007,Moustakas2013,Renzini2015}.
A widely accepted evolutionary scenario is that SFGs stop or reduce star formation and further transform into QGs
\citep[e.g.,][]{Faber2007,Peng2010}.

However, some QGs experience significant star formation bursts after a long period of low star formation, so-called ``rejuvenation''
\citep[e.g.,][]{Rowlands2018,Chauke2019},
which improves our understanding on galaxy evolution.
One rejuvenation event may form 10\% of the total mass \citep{Chauke2019,Tacchella2022}. 
Galaxies, especially massive ones, may experience multiple rejuvenation events \citep{Tanaka2024},
so the contributions of rejuvenation processes to the growth of galaxies are expected to be non-negligible.
Recently, \cite{Pandey2024} found that star formation exits in all early-type galaxies in the nearby Universe,
which suggests that rejuvenation is a constant process.
In addition, \cite{Mancini2019} argued that rejuvenations may be responsible for the bending of the star-forming main sequence at high-mass end.

Despite the importance of rejuvenation, the nature and mechanism of rejuvenation still remain unclear.
Based on a simulation, \cite{Kaviraj2009} demonstrate that gas accretion during gas-rich minor mergers
can trigger rejuvenation processes in early-type galaxies.
Other studies also suggest that these processes may further enable early-type galaxies turn into red spirals or S0 galaxies \citep[e.g.,][]{Diaz2018,Hao2019,Rathore2022}.
However, there is still a lack of direct observational evidences for the connection between rejuvenation and mergers.
Recently, \cite{Paudel2023} found an early-type dwarf galaxy is rejuvenating star formation through accreting gas from a nearby star-forming dwarf galaxy.
In this work, we report another rare case where a radio-loud active galactic nuclei (AGN), 3C 59, hosted by an early-type galaxy is rejuvenating star formation activity through ongoing minor mergers with its nearby satellite galaxies.

This paper is organized as follows. Section \ref{sec:identifysate} describes the identification of satellite galaxies for 3C 59.
Section \ref{sec:observationdata} introduces the observational data used to construct the broadband spectral energy distribution (SED).
Section \ref{sec:decomposition} describes the details about the image decomposition that obtains the structure properties of all the galaxies.
Section \ref{sec:SEDfit} introduces the detailed SED fitting processes.
We discuss the results in Section \ref{sec:resultdiscussion} and summarize them in Section \ref{sec:summary}.
Throughout this paper, we 
assume a flat cosmology with the following parameters:
$\Omega_{\rm m}=0.3$, $\Omega_{\Lambda}=0.7$, and $H_0=70\ \rm{km}\ \rm{s}^{-1}\ \rm{Mpc}^{-1}$.

% ########################      Figure 1     ##############################
\begin{figure*}[t!]
\centering
\includegraphics[width=0.85\linewidth, clip]{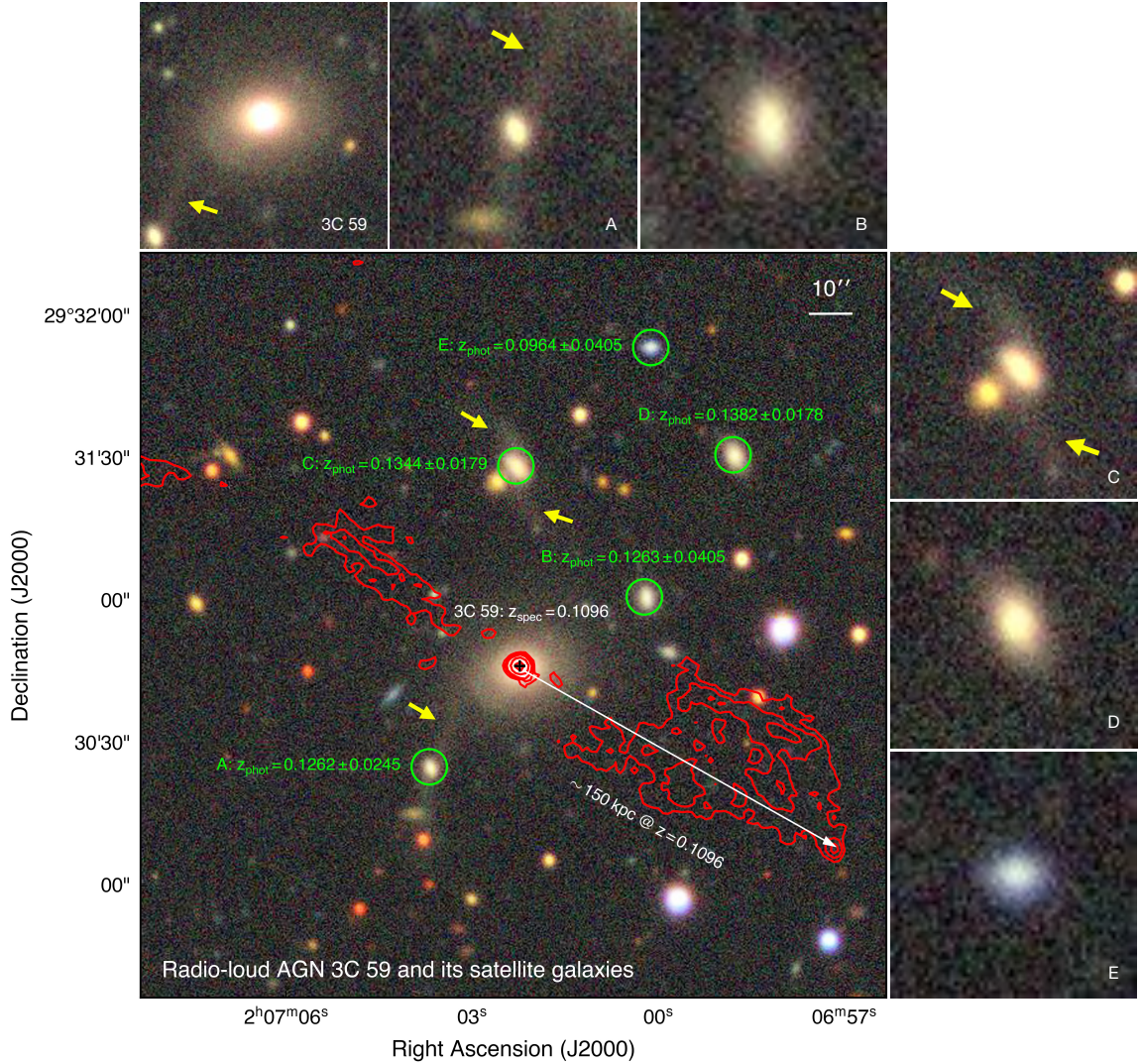}
\caption{Images for 3C 59 (the black cross) and its five satellite galaxies (A, B, C, D, and E; the green circles). 
The background shows an RGB image of DESI $g$, $r$, and $z$ filters.
The red contour represents the VLASS 3 GHz data,
and the contour levels are $5\sigma$, $8\sigma$, $12\sigma$, $30\sigma$, and $70\sigma$ ($\sigma \sim 0.18$ mJy beam$^{-1}$).
The white arrow represents the distance from 3C 59 to the edge of the radio lobe ($\sim$ 150 kpc at $z=0.1096$).
The yellow arrows highlight the tidal tails.
The six small panels show the individual galaxies zoomed in.
\label{fig:optimage}}
\end{figure*}

% #####################################################################
% ########################        Section 2       ##############################
% #####################################################################
\section{Identify satellite galaxies for 3C 59}
\label{sec:identifysate}
3C 59 is a Fanaroff-Riley Class II (FR II) radio AGN
with a large-scale radio jet extending to $\sim$ 150 kpc (see Figure \ref{fig:optimage}).
It has a spectroscopic redshift of $z_{\rm spec, 3C 59} = 0.1096$ \citep{Eracleous2004} and
a photometric redshift of $z_{\rm phot, 3C 59} = 0.1424 \pm 0.0161$ in Sloan Digital Sky Survey (SDSS) survey \citep{Ahumada2020}.
Satellite galaxies of 3C 59 were selected by the following criteria:
(1) the projected distance to 3C 59 should be less than 200 kpc
because $200$ kpc may be a threshold for the virial radius of 3C 59 dark matter halo 
(see details in Section \ref{sec:quenchingsatellite}); 
(2) the redshift difference between each satellite and 3C 59 
should be less than the redshift uncertainty of each satellite ($\delta_{z_{\rm phot, sate}}$): 
$|z_{\rm phot, sate} - z_{\rm spec, 3C 59}| \leq \delta_{z_{\rm phot, sate}}$ 
or $|z_{\rm phot, sate} - z_{\rm phot, 3C 59}| \leq \delta_{z_{\rm phot, sate}}$.
Most of the galaxies that lie around 3C 59 in projection only have photometric redshifts from SDSS survey.
Their photometric redshifts are estimated through the kd-tree nearest neighbor fit based on SDSS data at $u$, $g$, $r$, $i$, and $z$ filters \citep{Beck2016}.
This fit method is adopted via a spectroscopic training set comprised of 1 976 978 galaxies,
 which are relatively accurate with a standard deviation of 0.0205 \citep{Beck2016}.
Finally, we found five satellite galaxies
and named them A, B, C, D, and E 
in order of closest to farthest from 3C 59 (see the green circles in Figure \ref{fig:optimage}).
The projected distance from Satellites A, B, C, D, and E to 3C 59 is about 
56 kpc, 62 kpc, 84 kpc, 126 kpc, and 145 kpc, respectively.
Their photometric redshifts given by SDSS survey are 
$0.1262\pm 0.0245$ for Satellite A,
$0.1263 \pm 0.0405$ for Satellite B,
$0.1344 \pm 0.0179$ for Satellite C,
$0.1382 \pm 0.0178$ for Satellite D,
and $0.0964 \pm 0.0405$ for Satellite E (also see Figure \ref{fig:optimage}).

% #####################################################################
% ########################        Section 3       ##############################
% #####################################################################
\section{Observational data}
\label{sec:observationdata}

% ########################      Section 3.1     ##############################
\subsection{Optical and radio images}
\label{sec:opticalradioimage}

Optical images at $g$, $r$, and $z$ filters (see the RGB image in Figure \ref{fig:optimage})
were derived from the Dark Energy Spectroscopic Instrument (DESI) Legacy Imaging Surveys Date Release 9 (DR9). 
Depth of \textbf{imaging} decreases from $g$ filter to $z$ filter \citep{Dey2019}.
The 5$\sigma$ point source (AB) depths with single exposure\footnote{See details in \url{https://www.legacysurvey.org/dr9/description/}}
are $\sim$ 24.2 mag for $g$ filter, $\sim$ 23.9 mag for $r$ filter, and $\sim$ 22.7 mag for $z$ filter.
Therefore, in the following analysis, we mainly used the results obtained 
with $g$-filter image that has the highest data quality due to the deepest imaging.

The 3 GHz radio image (see the red contour in Figure \ref{fig:optimage}) was derived from the 
Very Large Array Sky Survey \citep[VLASS;][]{Lacy2020} Epoch 2.2. 

% ########################      Section 3.2     ##############################
\subsection{Data used for constructing a broadband SED of 3C~59}
\label{sec:3c59data}

To construct a broadband SED
with the Code Investigating GALaxy Emission \citep[\textsc{cigale};][]{Burgarella2005,Noll2009,Boquien2019,Yang2020,Yang2022} (see Section \ref{sec:3c59SED}),
we simultaneously utilized the multi-wavelength photometric data from 
the historical published catalogs 
and the optical spectrum from the Large Sky Area Multi-Object Fiber Spectroscopic Telescope (LAMOST).

{\it Multi-wavelength photometric data:} The multi-wavelength data from UV to radio bands were collected from the following multiple surveys:
Galaxy Evolution Explorer (GALEX), SDSS, Two Micron All-Sky Survey (2MASS), Wide-field Infrared Survey Explorer (WISE), Infrared Astronomical Satellite (IRAS), VLASS, and very long baseline interferometry (VLBI).
The cross-matching radius with each catalog is 1$''$. 
We refer readers to Appendix \ref{appendixsec:photodata} for detailed introduction about these surveys and data.
In Table \ref{tab:3c59data}, we summarize the information about the multi-wavelength data.  

{\it LAMOST optical spectrum (3700--9000~\AA):}
\textsc{cigale} cannot directly analyze spectrum data.
Therefore, we selected eight continuum ranges in the calibrated LAMOST spectrum 
(see the yellow regions in Figure \ref{fig:LAMOSTspectra}), 
and the average wavelength and flux of each range were utilized in the SED fitting. 
We refer readers to Appendix \ref{appendixsec:LAMOSTspectrum} for more details about the LAMOST spectrum.

% ########################      Section 3.3     ##############################
\subsection{Data used for constructing broadband SEDs of the five satellite galaxies}
\label{sec:compdata}

For the five satellite galaxies, we collected historical public data from SDSS, WISE, and 2MASS surveys.
The cross-matching radius with each catalog is $1''$.
The basic informations about these data 
are summarized in Table \ref{tab:compdata}.

For 3C 59 and all the satellite galaxies, we followed \cite{Schlafly2011} to make the Galactic extinction correction for all the observational data including photometric data and the LAMOST spectrum. In this process, we adopted the \cite{Fitzpatrick1999} reddening law with $R_V=3.1$ based on the dust map from \cite{Schlegel1998}.

% ########################      Figure 2     ##############################
\begin{figure*}[t!]
\centering
\includegraphics[width=0.83\linewidth, clip]{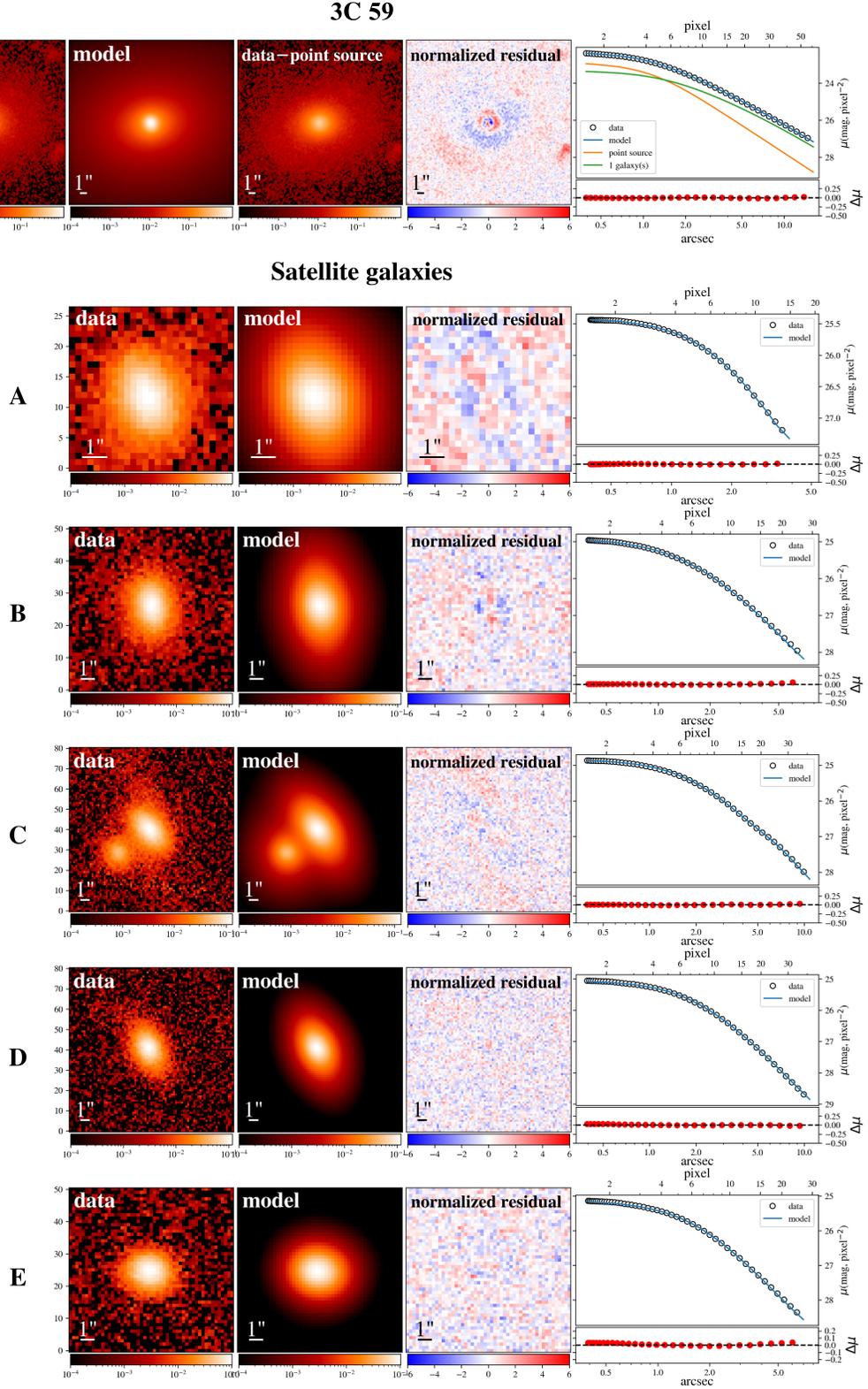}
\caption{2D decomposition of 3C 59 and the five satellite galaxies for the DESI $g$-band images with \textsc{galight}. 
The first row is shown for 3C 59, while the panels from left to right represent data, model, data minus point source (host galaxy only), normalized residual (data$-$model$/$noise), and surface brightness profile, respectively.
The second row to the sixth row are shown for Satellites A, B, C, D, and E, respectively,
while in each row, the panels from left to right represent data, model, normalized residual, and surface brightness profile, respectively.
\label{fig:compdecomposition}}
\end{figure*}

% #####################################################################
% ########################        Section 4       ##############################
% #####################################################################
\section{Two-dimensional DESI IMAGE DECOMPOSITION}
\label{sec:decomposition}

The image decomposition for DESI images at $g$, $r$, and $z$ filters 
were performed based on \textsc{galight} software \citep{Ding2021},
which is a Python code that provides various processing tools and two-dimensional (2D) profile fitting for astronomical data.
For 3C 59 and the five satellite galaxies, the best-fit parameters, such as the effective radius ($R_{\rm e}$), S$\acute{\rm e}$rsic index ($n_{\rm s\acute{\rm e}rsic}$), 
AGN fluxes, and reduced $\chi^2$ at each filter, are summarized in Table \ref{tab:structurepara}.

% ########################      Section 4.1     ##############################
\subsection{Image decomposition for 3C 59}
\label{sec:3c59decomp}
The AGN component was modeled using an empirical point-spread function (PSF) based on stars within the DESI survey area,
while the host galaxy component was modeled as a smooth S$\acute{\rm e}$rsic profile.
The 2D decomposition at $g$ filter is shown in Figure \ref{fig:compdecomposition}.
The decompositions at $r$ and $z$ filters show similar results,
so we do not exhibit them here.
The AGN fluxes at $g$, $r$, and $z$ filters
were determined using a combined PSF model based on two stars within a $3' \times 3'$ area centered on 3C 59,
which had the lowest reduced $\chi^2$ for the fit.
The uncertainties of the fluxes were estimated using five PSF models, each constructed from different stars. 
In addition,
simultaneously freeing the parameters of the PSF and S$\acute{\rm e}$rsic profiles 
results in a degeneracy among parameters. 
Thus, we made the 2D profile fitting with two tests: (1) $n_{\rm s\acute{\rm e}rsic}$ is fixed to 4 
that is a rough approximation of elliptical galaxies; 
(2) $n_{\rm s\acute{\rm e}rsic}$ is fixed to 1 that is an approximation of disk galaxies.
The fitting with $n_{\rm s\acute{\rm e}rsic}=4$ shows a better fit comparing to assuming $n_{\rm s\acute{\rm e}rsic} = 1$.
Therefore, we conclude that the host galaxy of 3C 59 shows an elliptical morphology. 

% ########################      Section 4.2     ##############################
\subsection{Image decomposition for satellite galaxies}
\label{sec:decompositionsatellite}
Each satellite galaxy was well modeled with only one smooth S$\acute{\rm e}$rsic profile.
The decomposition result for the $g$-filter image is shown in Figure \ref{fig:compdecomposition}. 
The decompositions for $r$- and $z$-filter images show similar results, which are not presented here.
All the satellite galaxies show a low $n_{\rm s\acute{\rm e}rsic}$ ($\sim 1$ or $\sim 2$), which indicates a disk (late-type) morphology.
However, these galaxies do not show significant spiral structures, which are consistent with S0 galaxy type (see Figure \ref{fig:compdecomposition}).

% ########################      Figure 3     ##############################
\begin{figure*}[t!]
\centering
\includegraphics[width=0.9\linewidth, clip]{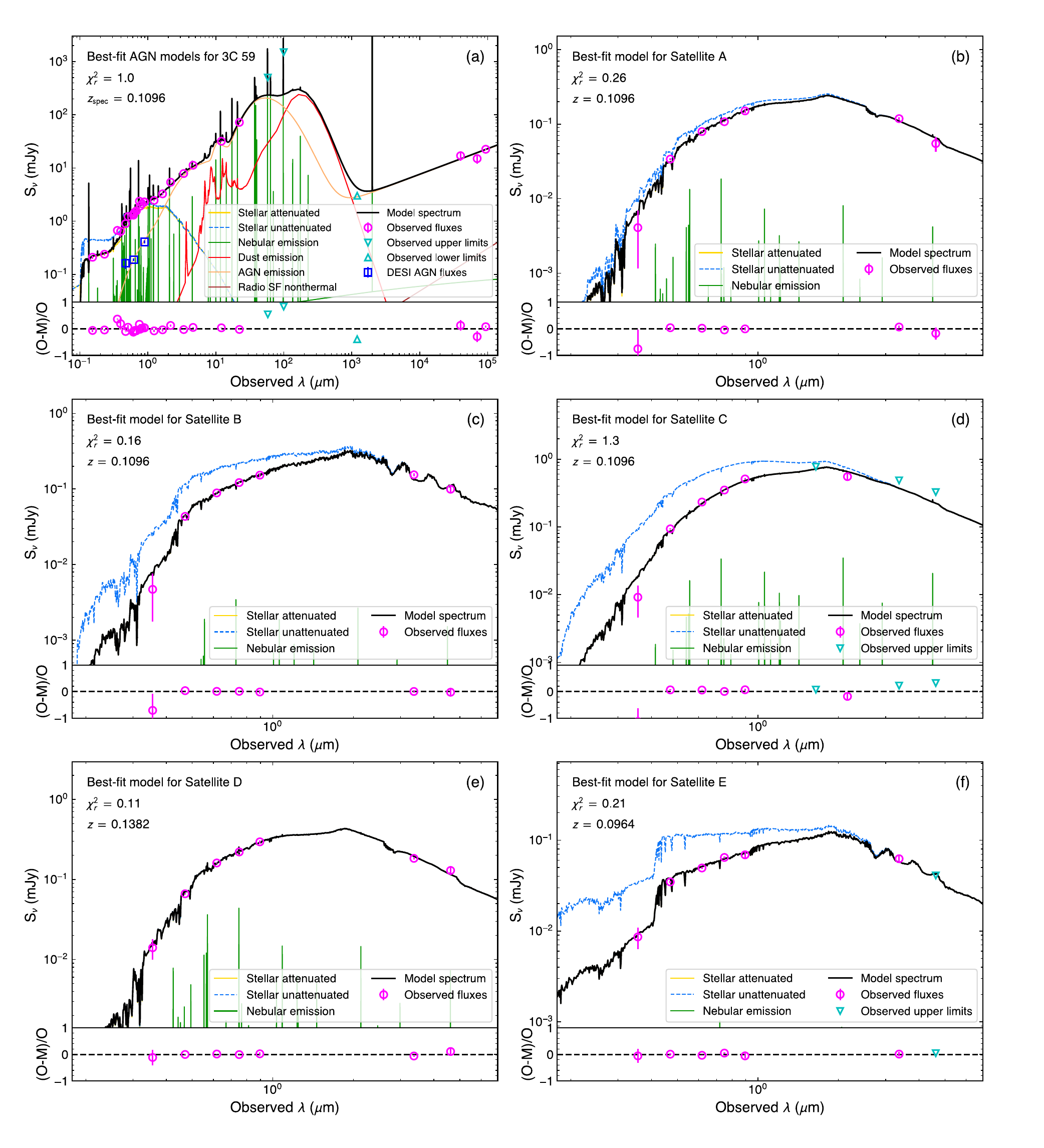}
\caption{Best-fit broadband SED model with \textsc{cigale} for 3C 59 (Panel a) and the five satellite galaxies (Panels b--f).
The details about the observational data used for SED fitting are introduced in Section \ref{sec:observationdata}, and shown in Tables \ref{tab:3c59data} and \ref{tab:compdata}.
The details about DESI AGN fluxes are introduced in Section \ref{sec:decomposition} and shown in Table \ref{tab:structurepara}.
\label{fig:totalSED}}
\end{figure*}

% ########################      Table 1     ##############################
   \begin{table*}[t!]
   \caption{Best-fit physical and structure properties for 3C 59 and the five satellite galaxies (A, B, C, D, and E) \label{tab:physicalpara}}
   \centering
   \footnotesize
   \setlength{\tabcolsep}{2pt}
   \begin{tabular}{cllcccccccccccc}
   \hline\hline
   Component & Parameter & Symbol & 3C~59 & A & B & C & D & E \\
   \hline
   & Redshift used in the fitting & $z_{\textsc{cigale}}$ & 0.1096 & 0.1096 & 0.1096 & 0.1096 & 0.1382 & 0.0964 \\
   \hline
   \multirow{4}{*}{Host galaxy} & Stellar mass & $M_\star$ ($10^{10}\ M_\odot$) & $4.37\pm 0.92$ & $0.80\pm 0.27$ & $0.57\pm 0.27$ & $3.02\pm 1.03$ & $2.29\pm 0.78$ & $0.10 \pm 0.04$ \\
    & Star formation rate & SFR ($M_\odot\ {\rm yr}^{-1}$) & $6.21\pm 0.71$ & $<0.21$ & $<1.09$ & $< 0.39$ & $<0.32$ & $<2.39$ \\
     & Dust attenuation & E(B-V) (mag) & $0.077\pm 0.004$ & $0.14 \pm 0.12$ & $0.28 \pm 0.16$ & $0.25 \pm 0.17$ & $0.10 \pm 0.10$ & $0.26 \pm 0.13$ \\
     & 8--1000 $\mu$m dust luminosity & $L_{\rm IR,dust}$  ($10^{37}$ W) & $2.42 \pm 0.12$ & {\nodata} & {\nodata} & {\nodata} & {\nodata} & {\nodata} \\
   \hline
   \multirow{6}{*}{AGN} & AGN dust reemitted luminosity & $L_{\rm IR, AGN}$ ($10^{37}$ W) & $9.81 \pm 0.66$ & {\nodata} & {\nodata} & {\nodata} & {\nodata} & {\nodata} \\
     & Fraction of $L_{\rm IR, AGN}$ to $L_{\rm IR, tot}$ & $F_{\rm AGN}$ & $0.80 \pm 0.11$ & {\nodata} & {\nodata} & {\nodata} & {\nodata} & {\nodata} \\
     & Polar dust attenuation & E(B-V) (mag) & $0.83 \pm 0.07$ & {\nodata} & {\nodata} & {\nodata} & {\nodata} & {\nodata} \\
     & AGN radio-loudness & $R_{\rm AGN}$ & $14 \pm 2$ & {\nodata} & {\nodata} & {\nodata} & {\nodata} & {\nodata} \\
     & Radio spectral index of AGN & $\alpha_{\rm AGN}$ & $0.41 \pm 0.12$ & {\nodata} & {\nodata} & {\nodata} & {\nodata} & {\nodata} \\
     & Viewing angle & $\theta_{\rm AGN}\ (^{\circ})$ & $23 \pm 14$ & {\nodata} & {\nodata} & {\nodata} & {\nodata} & {\nodata} \\
   \hline
   \end{tabular}
   \end{table*}

% #####################################################################
% ########################        Section 5       ##############################
% #####################################################################
\section{Broadband SED analysis}
\label{sec:SEDfit}

We constructed the broadband SED  
with \textsc{cigale}.
The \textsc{cigale} is a Python code that efficiently models the observational multi-wavelength data 
from X-ray to radio bands using a series of AGN/galaxies templates.
Based on the Bayesian-like approach, this code estimates many fundamental physical properties of host galaxies and AGN,
such as star formation rate (SFR), stellar mass ($M_{\star}$), dust luminosity, AGN contribution and other quantities.

% ########################      Section 5.1     ##############################
\subsection{Broadband SED fitting for 3C 59}
\label{sec:3c59SED}

Simultaneously fitting AGN and host galaxy components of 3C 59 cannot better constrain the various parameters.
Therefore, we firstly obtained the best-fit model for AGN and dust components,
and then fixed them in the entire broadband SED fitting.

{\it Fitting for the AGN and dust components:} 
We used the AGN module \citep{Stalevski2012,Stalevski2016}, the dust emission \citep{Dale2014}, and the radio module in \textsc{cigale} to simultaneously fit the AGN fluxes at DESI $g$, $r$, and $z$ filters (see details in Section \ref{sec:3c59decomp}),
and photometric data from WISE, VLA, and VLBI.
At the reference wavelength of these data, 
SED is dominated by AGN or correlated with dust emission.
In addition, we did not consider the minor contribution of the nebular emission to the radio continuum.
The model parameter settings in the fit are summarized in Table \ref{tab:3c59AGNSEDpara}.
The best-fit SED is shown in Figure \ref{fig:3C59AGNSED}, the best-fit model parameters are listed in Table \ref{tab:3c59SEDpara},
and the best-fit AGN properties are summarized in Table \ref{tab:physicalpara}.
We refer readers to Appendix \ref{sec:AGNproperties3C59} for detailed discussions about AGN properties of 3C 59.

{\it Fitting for the entire broadband SED:} 
In addition to AGN, dust, and radio modules whose model parameters were fixed to the best-fit values obtained above,
we used three additional modules to fit all the multi-wavelength data of 3C 59.
These three modules are the delayed $\tau$ star formation history (SFH), the \cite{Bruzual2003} simple stellar population model, 
and the \cite{Calzetti2000} galaxy dust attenuation.
These modules and their parameter settings are summarized in Table \ref{tab:3c59SEDpara},
and the best-fit results for the host galaxy properties are listed in Table \ref{tab:physicalpara}.
The best-fit SED is shown in Panel (a) of Figure \ref{fig:totalSED}.

% ########################      Section 5.2     ##############################
\subsection{SED fitting for satellite galaxies}

For each satellite galaxy, the galaxy templates include the following three modules:
the delayed $\tau$ SFH, the \cite{Bruzual2003} simple stellar population model, 
and the \cite{Calzetti2000} galaxy dust attenuation.
These modules and their parameter settings are summarized in Table \ref{tab:compSEDpara}.
As we mentioned in Section \ref{sec:identifysate},
photometric redshifts of SDSS surveys are estimated based on solely SDSS data.
In this work, we also included WISE (and 2MASS) data in addition to SDSS data.
Therefore, to find a suitable redshift value applied in the SED fitting, 
we fixed the redshift value of each satellite galaxy 
to SDSS photometric redshift and spectroscopic redshift of 3C 59, respectively.
We found that for Satellites A, B, and C, fixing to the spectroscopic redshift 
of 3C 59 provides a better fitting (lower reduced $\chi^2$) than fixing to the SDSS photometric redshifts,
while for Satellites D and E, the opposite is true.
Therefore, the redshift value applied in the SED fitting is 0.1096 for Satellite A/B/C, 0.1382 for Satellite D, and 0.0964 for Satellite E (see Table \ref{tab:physicalpara}).
The best-fit SEDs are shown in Figure \ref{fig:totalSED}.

% ########################      Figure 3     ##############################
\begin{figure*}[h!]
\center
\includegraphics[width=0.9\linewidth, clip]{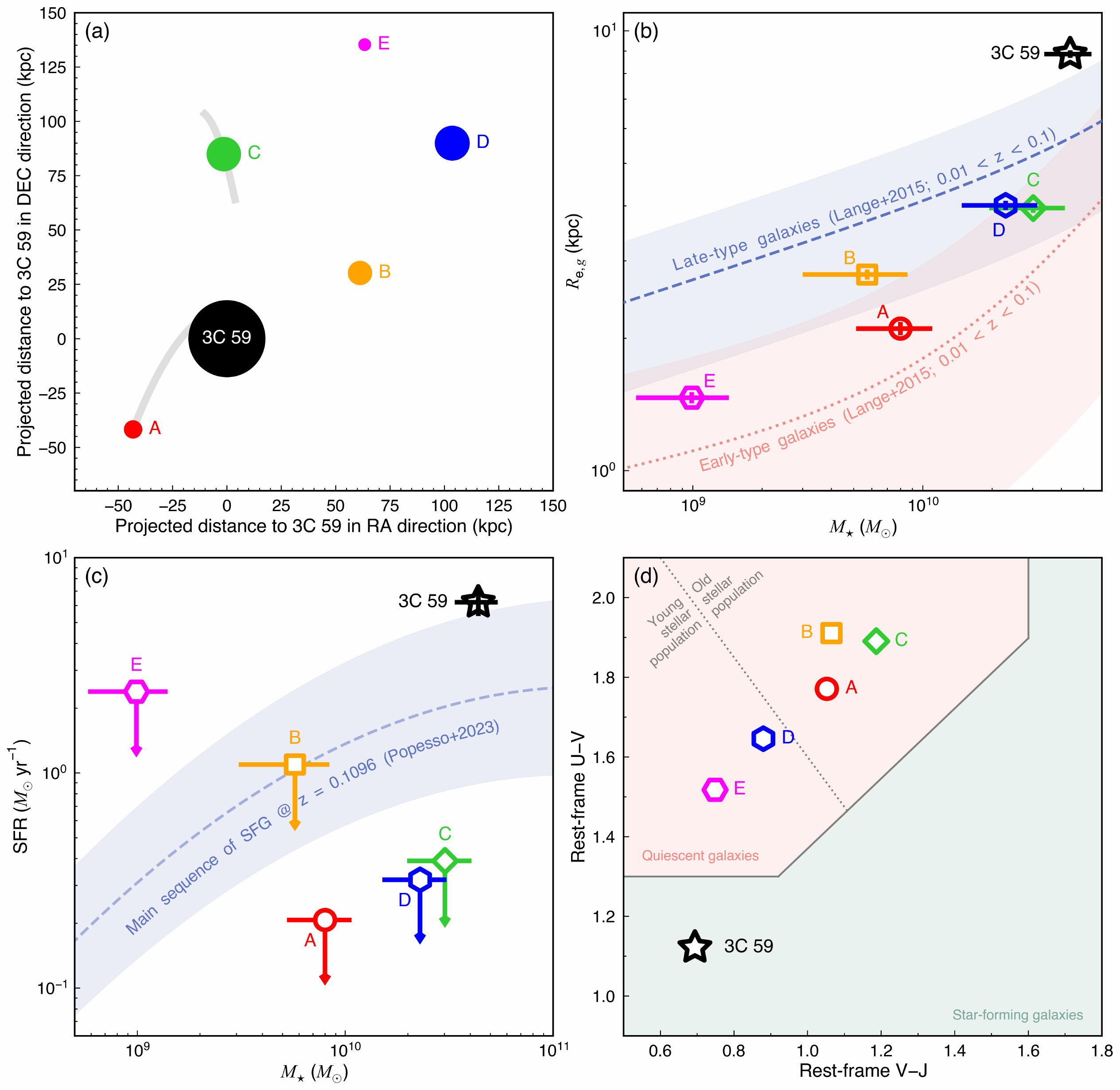}
\caption{Structure properties and star formation properties of 3C 59 and the five satellite galaxies (A, B, C, D, and E).
{\it Panel (a):} Projected distance of each satellite galaxy to 3C 59. The radius of each circle is equal to $2R_{\rm e,g}$ where $R_{\rm e,g}$ is the DESI $g$-filter effective radius (see Section \ref{sec:decomposition} and Table \ref{tab:structurepara}). The gray solid lines represent cartoon drawings of the tidal tails that correspond to the structures highlighted by the yellow arrows in Figure \ref{fig:optimage}.
{\it Panel (b):} $M_\star$--$R_{\rm e,g}$ relation. The $M_\star$ is the stellar mass (see Section \ref{sec:SEDfit} and Table \ref{tab:physicalpara}).
The blue and red regions show the $M_\star$-$R_{\rm e}$ relations for local late- and early-type galaxies \citep{Lange2015}, respectively.
{\it Panel (c):} $M_\star$--SFR relation. SFR is the star formation rate (see Section \ref{sec:SEDfit} and Table \ref{tab:physicalpara}).
{\it Panel (d):} Rest-frame UVJ diagram. Rest-frame U, V, and J magnitudes are estimated from the best-fit broadband SED (see Fig. \ref{fig:totalSED}).
The solid gray lines separating quiescent and star-forming galaxies are derived from \cite{Schreiber2015}.
The dotted gray line separating quiescent galaxies into young-stellar-population- and old-stellar-population-dominated galaxies are derived from \cite{Whitaker2013}.
\label{fig:scalingrelation}}
\end{figure*}

% #####################################################################
% ########################        Section 6       ##############################
% #####################################################################
\section{Results and discussion}
\label{sec:resultdiscussion}

% ########################      Section 6.1     ##############################
\subsection{Star formation rejuvenation in 3C 59}
\label{sec:rejuvenationSF3C59}
We combined two methods to decide the star-formation properties of each galaxy: (1) the star-forming main sequence; (2) rest-frame UVJ color diagram.
Throughout this work, we adopted the main sequence from \cite{Popesso2023}
 and UVJ selection criteria from \cite{Schreiber2015}.
 The main sequence from \cite{Popesso2023} is obtained based on a \cite{Kroupa2001} initial mass function (IMF), while we adopted a \cite{Chabrier2003} IMF in the SED fitting with \textsc{cigale}. Therefore, we multiplied $M_\star$ and SFR of the main sequence in \cite{Popesso2023} by a factor of 1.06 \citep{Speagle2014} to convert values from the \cite{Kroupa2001} IMF to the \cite{Chabrier2003} IMF.
3C 59 is located above the main sequence (see Panel c of Figure \ref{fig:scalingrelation}),
and resides in the star-forming region in the UVJ diagram (see Panel d of Figure \ref{fig:scalingrelation}).
Combining these two methods, we classified the host galaxy of 3C 59 as an SFG.

The reduced $\chi^2$ of 2D image decomposition (see Table \ref{tab:structurepara})
decreases from Satellite A to E (distance to 3C 59 from small to large, see Panel a of Figure \ref{fig:scalingrelation}),
which indicates that the closer the satellite galaxies are to 3C 59, the more significant the morphological disturbances are.
In addition, Satellites A and C exhibit significant tidal tails pointing towards 3C 59 (highlighted by yellow arrows in Figure \ref{fig:optimage} and gray lines in Panel a of Figure \ref{fig:scalingrelation}).
The effective radius of 3C 59 is much larger than the scaling relations for 
local early- and late-type galaxies \citep{Lange2015} (see Panel b of Figure \ref{fig:scalingrelation}).
This result indicates that 3C 59 is undergoing a dramatic size growth,
while many works have demonstrated that minor merger may be the main driver of size growth in early-type galaxies 
\citep[e.g.,][]{Naab2009,LopezSanjuan2012,Bedorf2013,Kaviraj2014,Oogi2016,Matharu2019}.
These pieces of observational evidences suggest that 3C 59 is undergoing a minor merger with its nearby satellite galaxies.

\begin{figure}[t!]
\center
\includegraphics[width=\linewidth, clip]{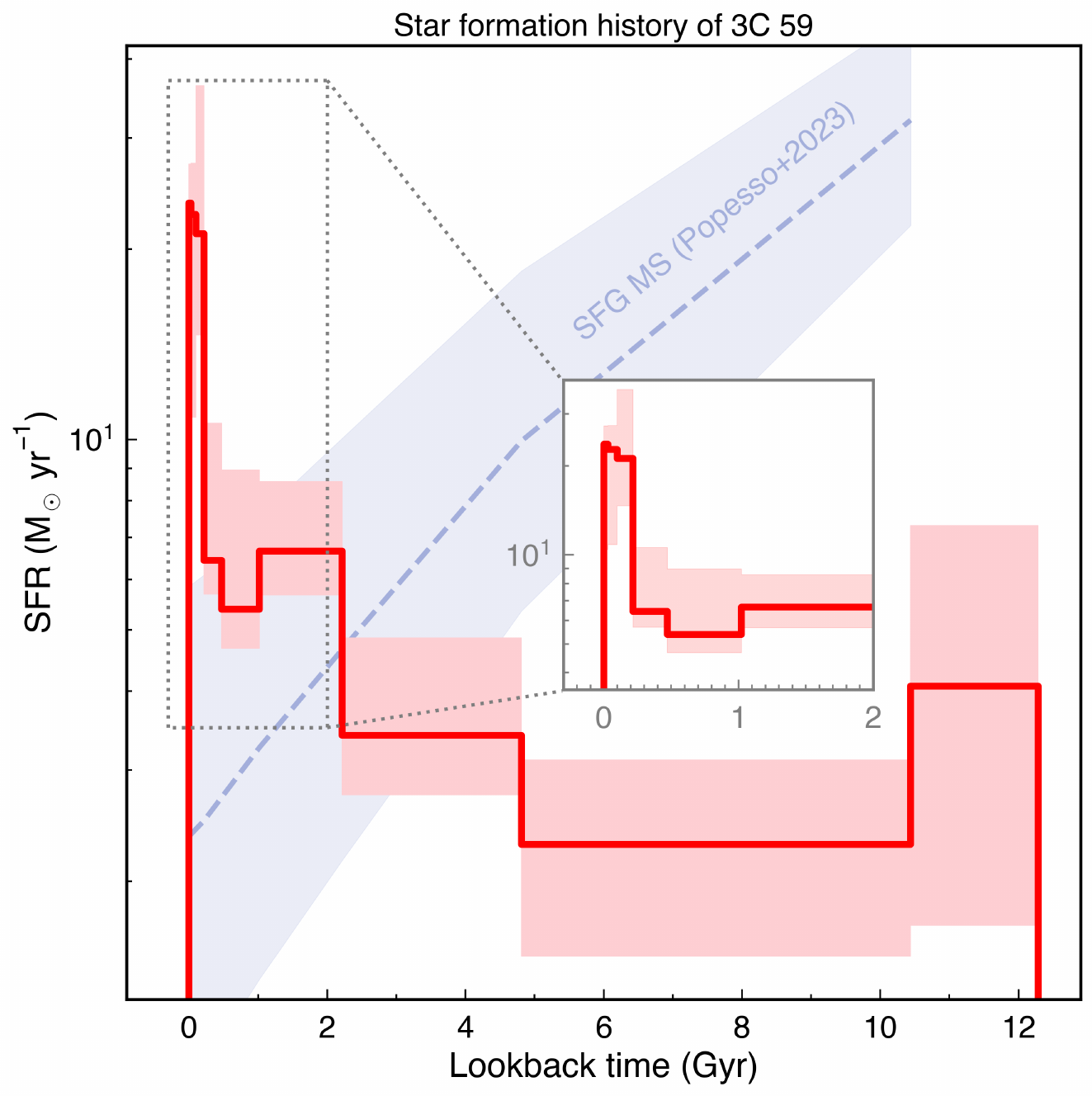}
\caption{The star formation history of 3C 59.
The red curve shows the best-fit star formation history of 3C 59 with \textsc{prospector},
while the red region shows the 1-$\sigma$ uncertainty.
The small grey panel shows the star formation history after zooming in recent 2 Gyr.
The dashed blue line represents the star-forming main sequence from \cite{Popesso2023},
while the blue region represents the 1-$\sigma$ uncertainty.
\label{fig:3C59SFH}}
\end{figure}

To investigate whether the minor merger affects star formation activities in 3C 59, we simultaneously fitted the multi-wavelength photometric data and LAMOST spectrum of 3C 59 with \textsc{prospector} \citep{Leja2017,Johnson2021}.
The \textsc{prospector} is a robust code that can provide a non-parametric SFH.
In the fit, we removed strong emission lines and subtracted AGN continuum model (obtained with \textsc{cigale}) from the observational data to exclude the bias brought by AGN activities.
We refer readers to Appendix \ref{sec:prospector3C59} for more details about the parameter setting in the fit.
The SFH of 3C 59 shows a significant rejuvenation of star formation activities within the past 500 Myr, before which remains rather quiescent for most of the cosmic time (see Figure \ref{fig:3C59SFH}). 
For the five satellite galaxies, we cannot well constrain their SFHs due to the lack of sufficient multi-wavelength data.

The ongoing minor mergers and star formation rejuvenation in 3C 59 indicate
that 3C 59 may be accreting cold gases from satellites to rejuvenate star formation activity and increase size.
Minor mergers may also trigger quasar activities and promote central black holes growth through enhanced gas supply \citep{WT2023}.

% ########################      Figure 4     ##############################

% ########################      Section 6.2     ##############################
\subsection{How the physical properties of 3C 59 affect star formation activities of satellite galaxies}
\label{sec:quenchingsatellite}
Due to the data limitation, we only obtained the upper limits of SFR for all the satellite galaxies, 
most of which are below the main sequence (see Panel c of Figure \ref{fig:scalingrelation}).
In the UVJ color diagram, all the five satellite galaxies are located at the quiescent region (see Panel d of Figure \ref{fig:scalingrelation}). 
Thus, we classified all the five satellite galaxies as QGs.
Next we discuss some possible mechanisms to explain the star formation quenching in these satellite galaxies.

\begin{itemize}
    \item[(1)] {\it Galaxy interaction strips gas in satellite galaxies.} The interaction between 3C 59 and satellite galaxies may result in a gas loss in satellite galaxies, and further reduce their star formation activities. 
Satellites A, B, and C show ongoing interaction features with 3C 59 but have reduced star formation much earlier (old stellar population; see Panel d of Figure \ref{fig:scalingrelation}).
It implies that their star formation quenching occurs earlier than the ongoing galaxy interaction.
As for Satellites D and E, the two farthest satellites, they do not show significant interaction features.
Their young stellar populations (see Panel d of Figure \ref{fig:scalingrelation}),
indicators for recent star formation quenching, also excludes the possibility
that they have already passed by 3C 59 and travel to larger distance with gas completely striped.
Therefore, gas stripping through galaxy interaction cannot fully explain the low star formation activities in all the satellite galaxies.

    \item[(2)] {\it Radio AGN feedback affect the satellites population.}
\cite{Shabala2011} found that satellites lying in the projected path of the FR II radio jets are more prone to have star formation quenching than those outside the path, while the satellites in the path of FR I radio jets show similar star formation properties to those outside the path.
3C 59 shows FR II radio jets, but all satellite galaxies are offset from the radio lobes.
It indicates that in this system, large-scale radio jets may not have a direct effect on the populations of these satellite galaxies, but an indirect way, such as heating the intergalactic medium to further reduce the star-formation in the satellites \citep[e.g.,][]{Shen2019},
still cannot be excluded.

    \item[(3)] {\it Massive dark matter halo prevents gas inflow.} 
Halo quenching is usually expected to operate in halo masses higher than $10^{12}\ M_\odot$,
where halo gas becomes shock-heated and gas supply to galaxies is further prevented \citep{Birnboim2003,Dekel2006}.
We adopted two methods to estimate the dark matter halo mass ($M_{\rm h}$) of 3C 59:
(i) $M_{\star}$-$M_{\rm h}$ relation\footnote{$M_\star/M_{\rm h} = 0.093\times [(M_{\rm h}/10^{11.77})^{-1.00}+(M_{\rm h}/10^{11.77})^{0.702}]^{-1}$} in \cite{Girelli2020}; (ii) $M_{\rm BH}$-$M_{\rm h}$ relation\footnote{$(M_{\rm BH}/10^8 M_\odot) = 0.10\ (M_{\rm h}/10^{12} M_\odot)^{1.65}$} in \cite{Ferrarese2002}, where $M_{\rm BH}$ is the black hole mass. The $M_{\rm BH}$ of 3C 59 is estimated to be $10^{8.912}\ M_\odot$ through reverberation mapping method \citep{Wu2004}. The method (i) obtained $M_{\rm h, \star} = 9.48^{+2.76}_{-2.18}\times 10^{11}\ M_\odot$, while for method (ii), $M_{\rm h, BH} = 1.44\times 10^{13}\ M_\odot$. The $M_{\rm h, BH}$ is 15 times larger than $M_{\rm h, \star}$. 
According to the halo virial radius ($R_{\rm vir}$) calculation\footnote{$R_{\rm vir} = 120\ (M_{\rm h}/10^{11} M_\odot)^{1/3}$ kpc} from \cite{Dekel2006}, $M_{\rm h, \star}$ and $M_{\rm h, BH}$ correspond to $R_{\rm vir,\star} = 254^{+23}_{-21}$ kpc and $R_{\rm vir,BH} = 629$ kpc, respectively.
According to \cite{Woo2015}, for halo mass of $M_{\rm h, \star}$, 10\% to 60\% satellites are quenched,
while for halo mass of $M_{\rm h, BH}$, 60\% to 100\% satellites are quenched.
Assuming that the halo mass is approximate to $M_{\rm h, BH}$ can fully explain the quiescent satellite fraction around 3C 59.
Both of the above two methods show that the halo mass of 3C 59 is almost higher than $10^{12}\ M_\odot$, 
so halo quenching may have a significant impact on populations of satellite galaxies.
In addition, ram-pressure stripping effect may also exist in this massive halo,
which may also affect the gas supply and star formations in satellite galaxies.

\end{itemize}

Overall, these results indicate that the massive halo combined with the enormous power from the large-scale radio jet of 3C 59 
may keep the halo hot, prevent gas cooling, and further reduce star formation in the satellite galaxies.

% #####################################################################
% ########################        Section 7       ##############################
% #####################################################################
\section{Summary and conclusions}
\label{sec:summary}
We witness a rare galaxy group system
where the central radio-loud AGN 3C 59 is hosted by an elliptical galaxy with recent star formation rejuvenation
and satellite galaxies are all quiescent disk galaxies.
The main conclusions are as follows:
\begin{itemize}
    \item[(1)] 3C 59 is a flat spectrum radio quasar with a large-scale radio jet extending to $\sim 150$ kpc.
    It is hosted by an elliptical galaxy (early-type galaxy) but undergoes a significant star formation rejuvenation within the past 500 Myr, before which remains rather quiescent for most of the cosmic time.
    \item[(2)] 3C 59 has five satellite galaxies within a projected distance of 200 kpc. Three nearest satellite galaxies show significant morphological disturbances and two of them exhibit significant tidal tails pointing towards 3C 59. In addition, 3C 59 has larger effective radius than both of local late- and early-type galaxies. These pieces of evidences indicate a minor merger between 3C 59 and nearby satellite galaxies, which may further trigger the star formation rejuvenation in 3C 59.
    Minor mergers may be also responsible for trigging quasar activities.
    \item[(3)] All the satellite galaxies are disk galaxies (late-type galaxies), 
    but do not have significant star formation activities.
It may be due to the possibility that the massive halo combined with the enormous power from the large-scale radio jet of 3C 59 keep the halo hot, prevent gas cooling, and further reduce star formation in these satellite galaxies.
    
\end{itemize}

% ###########################################################################
% ########################       Acknowledgments      ##############################
% ###########################################################################
\section{Acknowledgments}
We thank the anonymous referee for the constructive comments that improved this paper.
We thank Fei Li, Lingrui Lin, Yulong Gao, Can Xu, Longyue Chen, Luwenjia Zhou, Min Bao, and Zhengyi Chen for helpful discussion in framing this work.
This work was supported by National Natural Science Foundation
of China (Project No. 12173017 and Key Project No.
12141301), National Key R\&D Program of China
(2023YFA1605600), and the China Manned Space Project
(No. CMS-CSST-2021-A07).
Y.J.W. acknowledges supports by National Natural Science Foundation of China (Project No. 12403019)
and Jiangsu Natural Science Foundation (Project No. BK20241188).
%JST | Jiangsu Natural Science Foundation | Basic Research Program of Jiangsu Province
%T.W. acknowledges support by National Natural Science Foundation of China (Project No. 12173017,
%and Key Project No. 12141301), National Key R\&D
%Program of China (2023YFA1605600), and the China
%Manned Space Project (No. CMS-CSST-2021-A07).
The DESI Legacy Imaging Surveys consist of three individual and complementary projects: the Dark Energy Camera Legacy Survey (DECaLS), the Beijing-Arizona Sky Survey (BASS), and the Mayall z-band Legacy Survey (MzLS). DECaLS, BASS and MzLS together include data obtained, respectively, at the Blanco telescope, Cerro Tololo Inter-American Observatory, NSF’s NOIRLab; the Bok telescope, Steward Observatory, University of Arizona; and the Mayall telescope, Kitt Peak National Observatory, NOIRLab. NOIRLab is operated by the Association of Universities for Research in Astronomy (AURA) under a cooperative agreement with the National Science Foundation. Pipeline processing and analyses of the data were supported by NOIRLab and the Lawrence Berkeley National Laboratory (LBNL). Legacy Surveys also uses data products from the Near-Earth Object Wide-field Infrared Survey Explorer (NEOWISE), a project of the Jet Propulsion Laboratory/California Institute of Technology, funded by the National Aeronautics and Space Administration. Legacy Surveys was supported by: the Director, Office of Science, Office of High Energy Physics of the U.S. Department of Energy; the National Energy Research Scientific Computing Center, a DOE Office of Science User Facility; the U.S. National Science Foundation, Division of Astronomical Sciences; the National Astronomical Observatories of China, the Chinese Academy of Sciences and the Chinese National Natural Science Foundation. LBNL is managed by the Regents of the University of California under contract to the U.S. Department of Energy. The complete acknowledgments can be found at \url{https://www.legacysurvey.org/acknowledgment/}.
Guoshoujing Telescope (the Large Sky Area Multi-Object Fiber Spectroscopic Telescope LAMOST) is a National Major Scientific Project built by the Chinese Academy of Sciences. Funding for the project has been provided by the National Development and Reform Commission. LAMOST is operated and managed by the National Astronomical Observatories, Chinese Academy of Sciences.
This research has made use of the CIRADA cutout service at URL cutouts.cirada.ca, operated by the Canadian Initiative for Radio Astronomy Data Analysis (CIRADA). CIRADA is funded by a grant from the Canada Foundation for Innovation 2017 Innovation Fund (Project 35999), as well as by the Provinces of Ontario, British Columbia, Alberta, Manitoba and Quebec, in collaboration with the National Research Council of Canada, the US National Radio Astronomy Observatory and Australia’s Commonwealth Scientific and Industrial Research Organisation.

\vspace{5mm}
\facilities{GALEX (FUV and NUV), SDSS ($u$, $g$, $r$, $i$, and $z$), LAMOST, WISE (W1, W2, W3, and W4), IRAS, VLA (3 GHz), VLBI (4.3 GHz and 7.6 GHz)
}


\software{astropy \citep{astropy2013,astropy2018},  
         CIGALE \citep{Burgarella2005, Noll2009, Boquien2019, Yang2020, Yang2022},
         GaLight \citep{Ding2021}
          }



\appendix

% ######################################################################
% ########################        Appendix A      ##############################
% ######################################################################
\section{Multi-wavelength photometric data}
\label{appendixsec:photodata}
The observational data at FUV and NUV filters of the Galaxy Evolution Explorer (GALEX) were derived from GALEX's All-Sky Imaging Survey \citep{Morrissey2007}.
The angular resolution is 4.2$''$ for FUV filter and 5.3$''$ for NUV filter \citep{Morrissey2007}.
The 5-$\sigma$ depth at FUV and NUV filters are 39.8 $\mu$Jy and 17.4 $\mu$Jy, respectively \citep{Morrissey2007,Bianchi2011}.

The observed fluxes at $u$, $g$, $r$, $i$, and $z$ bands of Sloan Digital Sky Survey (SDSS)
were collected from the 16th data release of SDSS \citep{Ahumada2020}.
A typical angular resolution of SDSS is about $1.2''$.
The 5-$\sigma$ depth at $u$, $g$, $r$, $i$, and $z$ bands are 5.0 $\mu$Jy, 2.0 $\mu$Jy, 3.0 $\mu$Jy, 4.8 $\mu$Jy, and 18.8 $\mu$Jy, respectively\footnote{See details in \url{https://www.sdss4.org/dr16/imaging/other_info/}}.

The Two Micron All-Sky Survey (2MASS) data in J (1.25 $\mu$m), H (1.65 $\mu$m), and Ks (2.17 $\mu$m) filters were derived from \cite{Cutri2003}.
The angular resolutions for these three filters are about 4.0$''$\footnote{See details in \url{https://www.ipac.caltech.edu/2mass/overview/about2mass.html}}.
The 5-$\sigma$ depth at J, H, and Ks filters are 0.39 mJy, 0.47 mJy, and 0.85 mJy, respectively \citep{Cutri2003}.

The Wide-field Infrared Survey Explorer (WISE) data in W1 (3.4 $\mu$m), W2 (4.6 $\mu$m), W3 (12 $\mu$m), and W4 (22 $\mu$m) bandpasses were collected from the ALLWISE program \citep{Cutri2014}.
Angular resolution is 6.1$''$ for W1, 6.4$''$ for W2, 6.5$''$ for W3, and 12.0$''$ for W4 \citep{Wright2010}.
The 5-$\sigma$ depth in W1, W2, W3, and W4 bandpasses are 0.08 mJy, 0.11 mJy, 1.0 mJy, and 6.0 mJy, respectively \citep{Wright2010}.

The Infrared Astronomical Satellite (IRAS) mission observed 3C 59 at 12 $\mu$m, 25 $\mu$m, 60 $\mu$m, 100 $\mu$m through IRAS Sky Survey Atlas  \citep{irsa10}, but did not find any counterpart. It might be due to the fact that the source fluxes are below the detection limits that is about 500 mJy at 12--60 $\mu$m, and about 1500 mJy at 100 $\mu$m\footnote{See details in \url{https://lambda.gsfc.nasa.gov/product/iras/}}.

The lower limit of flux at 250 GHz was obtained with the Max-Planck Millimeter Bolometer (MAMBO) at the IRAM 30m telescope, which was derived from \cite{Bemmel2001}.
The \textsc{cigale} cannot utilize lower limits of fluxes, so we only use this lower limit as a reference for whether the SED fitting at sub-millimeter is reliable.

The VLASS 3 GHz photometric data was derived from \cite{Gordon2021}.
The angular resolution is about $2.5''$.
The median rms for images is 128 $\mu$Jy beam$^{-1}$ \citep{Gordon2021}.

The 4.3 GHz and 7.6 GHz data detected by the very long baseline interferometry (VLBI) were derived from \cite{Petrov2021}.
Both of these two bands have an angular resolution below 0.1$''$.
The characteristic image sensitivities\footnote{See details in \url{https://science.nrao.edu/facilities/vlba/docs/manuals/oss/bands-perf}} at 4.3 GHz and 7.6 GHz are $\sim$ 5 mJy beam$^{-1}$ and $\sim$ 8--10 mJy beam$^{-1}$, respectively.

\setcounter{table}{0}
\renewcommand{\thetable}{A\arabic{table}}

\setcounter{figure}{0}
\renewcommand{\thefigure}{A\arabic{figure}}

% ########################      Table A1     ##############################
   \begin{table*}[h!]
   \caption{Multi-wavelength data used for constructing the broadband SED of 3C 59. \label{tab:3c59data}}
   \centering
   \small
   \begin{tabular}{cccccccc}
   \hline\hline
 Observatory/survey & Observation date & Reference  & Filter/band  & ${\lambda_{\rm ref}}$ (\AA) $^1$ & Flux (mJy) $^2$ \\
   \hline
GALEX & 2007-9-10 & \cite{Bianchi2011} & FUV & 1535 & $0.210 \pm 0.014$ \\
             &                  &                               & NUV & 2301 & $0.239 \pm 0.010$ \\
             \hline
SDSS & 2009-10-17 & \cite{Ahumada2020} & u & 3557 & $0.670 \pm 0.007$ \\
           &                    &                                   & g & 4703 & $0.899 \pm 0.003$ \\
           &                    &                                   & r & 6176 & $1.376 \pm 0.005$ \\
           &                    &                                   & i & 7490 & $2.375 \pm 0.009$ \\
           &                    &                                   & z & 8947 & $2.299 \pm 0.015$ \\
             \hline
LAMOST $^3$ & 2016-01-12 & LAMOST DR8  & band-1 & 3973 & $0.644 \pm 0.142$ \\
               &                     &                          & band-2 & 5096 & $1.219 \pm 0.110$ \\
               &                     &                          & band-3 & 6126 & $1.297 \pm 0.074$ \\
               &                     &                          & band-4 & 6399 & $1.467 \pm 0.068$ \\
               &                     &                          & band-5 & 6763 & $1.557 \pm 0.103$ \\
               &                     &                          & band-6 & 7694 & $1.918 \pm 0.114$ \\
               &                     &                          & band-7 & 8197 & $2.114 \pm 0.155$ \\
               &                     &                          & band-8 & 8794 & $2.295 \pm 0.220$ \\
             \hline
2MASS & 1999-11-24 & \cite{Cutri2003} & J                   & 12350 & $2.454\pm 0.108$ \\
             &                    &                             & H                   & 16620 & $3.265\pm 0.192$ \\
             &                    &                             & $K_{\rm s}$ & 21590 & $5.395\pm 0.194$ \\
             \hline
WISE    & 2013           & \cite{Cutri2014} & W1 & 33526 & $7.86\pm 0.17$ \\
             &                    &                            & W2 & 46028 & $11.25\pm 0.23$ \\
             &                    &                            & W3 & 115608 & $31.83\pm 0.50$ \\
             &                    &                            & W4 & 220883 & $72.06\pm 1.92$ \\
             \hline
IRAS $^4$ & 1983  &  \cite{irsa10} & IRAS60 & $6 \times 10^5$ & $< 500$ \\
         &                      &                                       & IRAS100 & $10^6$ & $< 1500$ \\
             \hline
IRAM & 1999 December           & \cite{Bemmel2001} & 250 GHz &    $1.2\times10^7$        &       $> 3$ \\
             \hline
VLASS & 2019-04-07           & \cite{Gordon2021} & 3 GHz &    $10^9$        &       $22.49\pm 0.80$ \\
             \hline
VLBI $^4$ &  2015--2016  &  \cite{Petrov2021}  & 4.3 GHz & $7 \times 10^8$  &  15.0 \\
         &                      &                               &  7.6 GHz & $4 \times 10^8$  &  17.0 \\
  \hline
   \end{tabular}
   \tablecomments{$^1$ Reference wavelength of filters are provided by \href{http://svo2.cab.inta-csic.es/theory/fps/index.php?mode=browse&gname=WISE&asttype=}{Filter Profile Service} except for the LAMOST spectrum. 
$^2$ The Galactic extinction corrections had been made for all the fluxes (see details in Section \ref{sec:3c59data}).
$^3$ From the calibrated LAMOST spectrum, we select 8 continuum ranges (band-1 to band-8 here; see the yellow regions in Figure \ref{fig:LAMOSTspectra}), and obtained the average wavelength and average flux of each range. 
$^4$ The flux uncertainties are not given by \cite{Petrov2021}, so we took 10\% of fluxes as the flux uncertainties in the broadband SED fitting.
   }
   \end{table*}

% ########################      Table A2     ##############################
   \begin{table*}[h!]
   \caption{Multi-wavelength data used for constructing the broadband SEDs of the five satellite galaxies (A, B, C, D, and E). \label{tab:compdata}}
   \centering
   \scriptsize
   \setlength{\tabcolsep}{3pt}
   \begin{tabular}{cccccccccccc}
   \hline\hline
 Observatory & Obs. date & Reference & Filter   & $\lambda_{\rm ref}$ ($\AA$) $^1$ & \multicolumn{5}{c}{Flux (mJy)} \\
  \cmidrule(lr){6-10}
   &   &  &  &  & A & B & C & D & E \\
\hline
SDSS & 2009-10-17 & \cite{Ahumada2020} & u & 3557 & $0.004 \pm 0.003$ & $0.005 \pm 0.003$ & $0.009 \pm 0.005$ & $0.014 \pm 0.004$ & $0.009 \pm 0.002$ \\
           &                    &                                  & g & 4703 & $0.034 \pm 0.001$ & $0.043 \pm 0.001$ & $0.094 \pm 0.002$ & $0.066 \pm 0.001$ & $0.035 \pm 0.001$ \\
           &                    &                                  &  r  & 6176 & $0.079 \pm 0.002$ & $0.089 \pm 0.002$  & $0.233 \pm 0.003$ & $0.161 \pm 0.003$ & $0.049 \pm 0.001$ \\
           &                    &                                  & i  & 7490 & $0.108 \pm 0.002$ & $0.122 \pm 0.003$ & $0.350 \pm 0.004$ & $0.220 \pm 0.003$ & $0.064 \pm 0.002$ \\
           &                    &                                  & z & 8947 & $0.150 \pm 0.010$ & $0.152 \pm 0.010$ & $0.510 \pm 0.016$ & $0.295 \pm 0.014$ & $0.069 \pm 0.007$ \\
             \hline
2MASS & 1999-11-24 & \cite{Cutri2003}   & H                   & 16620 & {\nodata} & {\nodata} & $< 0.772$ & {\nodata} & {\nodata} \\
             &                    &                             & $K_{\rm s}$ & 21590 & {\nodata} & {\nodata} & $0.552\pm 0.069$ & {\nodata} & {\nodata} \\
             \hline
WISE    & 2013           &            \cite{Cutri2014}               & W1 & 33526 & $0.118 \pm 0.006$ & $0.154 \pm 0.006$ & $< 0.486$ & $0.183 \pm 0.007$ & $0.062 \pm 0.005$ \\
             &                    &                                                    & W2 & 46028 & $0.055 \pm 0.012$ & $0.099 \pm 0.011$ & $< 0.327$ & $0.130 \pm 0.016$ & $< 0.041$ \\
  \hline
   \end{tabular}
   \tablecomments{$^1$ Reference wavelength of filters are provided by \href{http://svo2.cab.inta-csic.es/theory/fps/index.php?mode=browse&gname=WISE&asttype=}{Filter Profile Service}.
The projected distance between Satellite C and its projected neighbor galaxy is about $5.5''$, 
which is lower than the angular resolution of WISE ($> 6.0''$, see Appendix \ref{appendixsec:photodata}). 
In \cite{Cutri2014}, Satellite C and its neighbor belong to one WISE-detected source, so here we used the WISE fluxes as upper limits for Satellite C.
}
   \end{table*}

% ######################################################################
% ########################        Appendix B      ##############################
% ######################################################################
\newpage
\section{Optical spectrum of 3C 59 from LAMOST}
\label{appendixsec:LAMOSTspectrum}
The one-dimensional (1D) LAMOST spectrum of 3C~59 was downloaded from the 
LAMOST DATA RELEASE 8 (DR8) low-resolution spectra website\footnote{See the LAMOST DR8 website: \url{http://www.lamost.org/dr8/}}.
LAMOST spectra cover a wavelength range of 3700--9000~{\AA} with a spectral resolution of $R\sim$1000--2000.
Each spectrum consists of a blue (3700--5900~\AA) and 
a red spectrograph arms (5700--9000~\AA)\footnote{See the LAMOST website: \url{http://www.lamost.org/public/node/119?locale=en}}.
LAMOST only makes relative flux calibration rather than accurate absolute flux calibration,
so we utilized SDSS multi-band ($g$, $r$, $z$) photometric data to recalibrate LAMOST red and blue spectra following \cite{Jin2023} (See Figure \ref{fig:LAMOSTspectra}).
We adopted the $\chi^2$ minimum method to match the released LAMOST spectrum that only has the relative flux without units
to the SDSS photometric data. Here we only used $g$, $r$, and $z$ bands of SDSS data to make flux calibration without including $u$ and $i$ filters, because $u$ band is out of the LAMOST spectrum coverage and $i$ band covers the wavelength range of H$\alpha$ emission lines (See Figure \ref{fig:LAMOSTspectra}).
During the re-calibration, we assumed that the spectral shape in the blue and red channels are not changed.

\setcounter{table}{0}
\renewcommand{\thetable}{B\arabic{table}}

\setcounter{figure}{0}
\renewcommand{\thefigure}{B\arabic{figure}}

% ########################      Figure B1     ##############################
\begin{figure*}[h!]
\center
\includegraphics[width=\linewidth, clip]{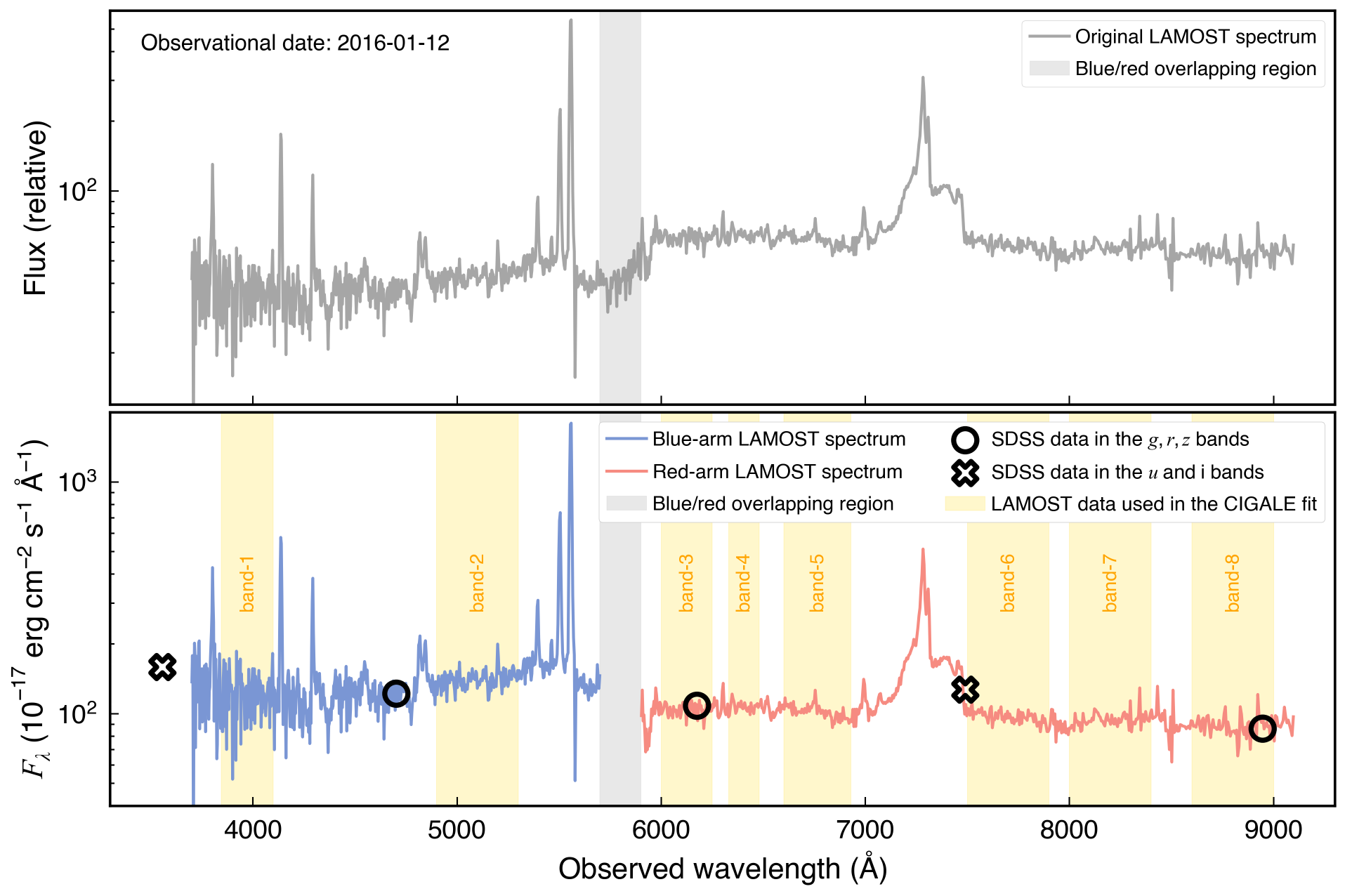}
\caption{Absolute flux calibration for the LAMOST spectrum of 3C 59. 
{\it Top panel:} the original LAMOST spectrum only with the relative flux calibration.
{\it Bottom panel:} the LAMOST spectrum after the absolute flux calibration. The blue and red curves represent the blue- and red-arm LAMOST spectra, respectively. The grey curve represents the overlapping region between the blue- and red-arm LAMOST spectrum (5700--5900 {\AA}). The empty black circles represent the flux densities in the SDSS $g$, $r$, and $z$ bands that are used in the flux calibration, while the empty black x-shape-symbols represent the flux densities in the SDSS $u$ and $i$ bands that are not used in the calibration (see details in Appendix \ref{appendixsec:LAMOSTspectrum}). The yellow regions represent the spectrum data that are used in the broadband SED fitting (see detailed values in Table \ref{tab:3c59data}).
\label{fig:LAMOSTspectra}}
\end{figure*}

% ######################################################################
% ########################        Appendix C      ##############################
% ######################################################################
\newpage
\section{2D image decomposition}

\setcounter{table}{0}
\renewcommand{\thetable}{C\arabic{table}}

\setcounter{figure}{0}
\renewcommand{\thefigure}{C\arabic{figure}}

% ########################      Table C1     ##############################
   \begin{table*}[h!]
   \caption{Best-fit structure properties for 3C 59 and the five satellite galaxies (A, B, C, D, and E) with \textsc{galight} \label{tab:structurepara}}
   \centering
   \begin{tabular}{clccccccccccccc}
   \hline\hline
   Parameter & Symbol & 3C~59 & A & B & C & D & E \\
   \hline
    & & & \multicolumn{2}{c}{$g$-filter} & & & \\
   \hline
   Effective radius & $R_{{\rm e,}g}$ (kpc) & $8.86\pm 0.07$ & $2.10 \pm 0.05$ & $2.79 \pm 0.04$ & $3.96 \pm 0.05$ & $4.01 \pm 0.05$ & $1.47 \pm 0.03$ \\
   S$\acute{\rm e}$rsic index & $n_{{\rm s\acute{\rm e}rsic,}g}$ & $4.0^*$ & $1.61 \pm 0.13$ & $1.68 \pm 0.07$ & $1.26 \pm 0.04$ & $1.31 \pm 0.04$ & $0.78 \pm 0.06$ \\
   AGN fluxes & $S_{{\rm AGN,}g}$ (mJy) & $0.16 \pm 0.04$ & {\nodata} & {\nodata} & {\nodata} & {\nodata} & {\nodata} \\
   Reduced $\chi^2$ & $\chi^2_{{\rm r}}$ & 1.14 & 1.22 & 0.98 & 0.90 & 0.83 & 0.85 \\
   \hline
    & & & \multicolumn{2}{c}{$r$-filter} & & & \\
   \hline
   Effective radius & $R_{{\rm e,}r}$ (kpc) & $9.00 \pm 0.06$ & $1.66 \pm 0.03$ & $2.55 \pm 0.04$ & $3.47 \pm 0.04$ & $4.01 \pm 0.05$ & $1.00 \pm 0.03$ \\
   S$\acute{\rm e}$rsic index & $n_{{\rm s\acute{\rm e}rsic,}r}$ & $4.0^*$ & $1.66 \pm 0.16$ & $1.96 \pm 0.12$ & $1.64 \pm 0.05$ & $1.31 \pm 0.04$ & $1.56 \pm 0.24$\\
   AGN fluxes & $S_{{\rm AGN,}r}$ (mJy) & $0.19 \pm 0.01$ & {\nodata} & {\nodata} & {\nodata} & {\nodata} & {\nodata} \\
   Reduced $\chi^2$ & $\chi^2_{{\rm r}}$ & 1.12 & 1.58 & 1.05 & 0.95 & 0.83 & 0.76 \\
   \hline
    & & & \multicolumn{2}{c}{$z$-filter} & & & \\
   \hline
   Effective radius & $R_{{\rm e,}z}$ (kpc) & $8.85 \pm 0.05$ & $2.60 \pm 0.21$ & $3.08 \pm 0.07$ & $3.52 \pm 0.04$ & $4.02 \pm 0.07$ & $1.16 \pm 0.05$ \\
   S$\acute{\rm e}$rsic index & $n_{{\rm s\acute{\rm e}rsic,}z}$ & $4.0^*$ & $5.25^{+0.95}_{-0.76}$ & $2.18 \pm 0.12$ & $1.80 \pm 0.04$ & $2.18 \pm 0.08$ & $2.33 \pm 0.36$ \\
   AGN fluxes & $S_{{\rm AGN,}z}$ (mJy) & $0.41 \pm 0.03$ & {\nodata} & {\nodata} & {\nodata} & {\nodata} & {\nodata} \\
   Reduced $\chi^2$ & $\chi^2_{{\rm r}}$ & 1.08 & 1.26 & 1.07 & 1.01 & 0.85 & 0.78 \\
   \hline
   \end{tabular}
   \tablecomments{$^{*}$ The S$\acute{\rm e}$rsic index of 3C 59 is fixed to 4.0 during the image decomposition (see details in Section \ref{sec:3c59decomp}).
   }
   \end{table*}

% ######################################################################
% ########################        Appendix D      ##############################
% ######################################################################
\newpage
\section{Model parameters for broadband SED fitting with \textsc{cigale}}
\setcounter{table}{0}
\renewcommand{\thetable}{D\arabic{table}}

\setcounter{figure}{0}
\renewcommand{\thefigure}{D\arabic{figure}}

% ########################      Table D1     ##############################
   \begin{table*}[h!]
   \caption{Model parameter settings in the SED fitting with only AGN and dust models for 3C 59  \label{tab:3c59AGNSEDpara}}
   \centering
   \small
   \setlength{\tabcolsep}{3pt}
   \begin{tabular}{clcccccc}
   \hline\hline
   Module & Parameter & Symbol & Values \\
   \hline
Galactic dust emission & \multirow{2}{*}{Slope in $dM_{\rm{dust}} \propto U^{-\alpha}dU$} & \multirow{2}{*}{$\alpha$} & \multirow{2}{*}{1.5 to 3.5 (step 0.125)} \\
\citep{Dale2014} & & & \\
\hline
\multirow{2}{*}{AGN (UV-to-IR)} & AGN contribution to IR luminosity & $\rm{frac}_{\rm{AGN}}$ & 0.05, 0.1 to 0.95 (step 0.05), 0.99 \\           & Viewing angle $^*$ & $\theta$ & $0^{\circ}$, $10^{\circ}$, $20^{\circ}$, $30^{\circ}$, $40^{\circ}$, $50^{\circ}$ \\
 \citep{Stalevski2012,Stalevski2016} & Polar-dust color excess & $E(B-V)_{\rm{PD}}$ (mag) & 0.1 to 1.5 (step 0.1) \\
\hline
\multirow{3}{*}{Radio} & SF radio-IR correlation parameter & $q_{\rm{IR}}$ & 2.58$^{\dagger}$ \\
  & Radio-loudness parameter & $R_{\rm{AGN}}$ & 2 to 50 (step 2)  \\
  & Spectral index of AGN radio emission  & $\alpha_{\rm AGN}$ & 0.20 to 0.60 (step 0.01) \\
\hline
   \end{tabular}
   \tablecomments{$^{\dagger}$ The star-formation radio-IR correlation parameter ($q_{\rm IR}$) is fixed to the default value of \textsc{cigale}. $^{*}$ 3C 59 exhibits double-peaked broad emission lines \citep{Eracleous1994,Restrepo2022}, which indicates a face-on orientation, so the viewing angle is set to be between $0^{\circ}$ and $50^{\circ}$.
}
   \end{table*}

% ########################      Figure D1     ##############################
\begin{figure}[h!]
\center
\includegraphics[width=\linewidth, clip]{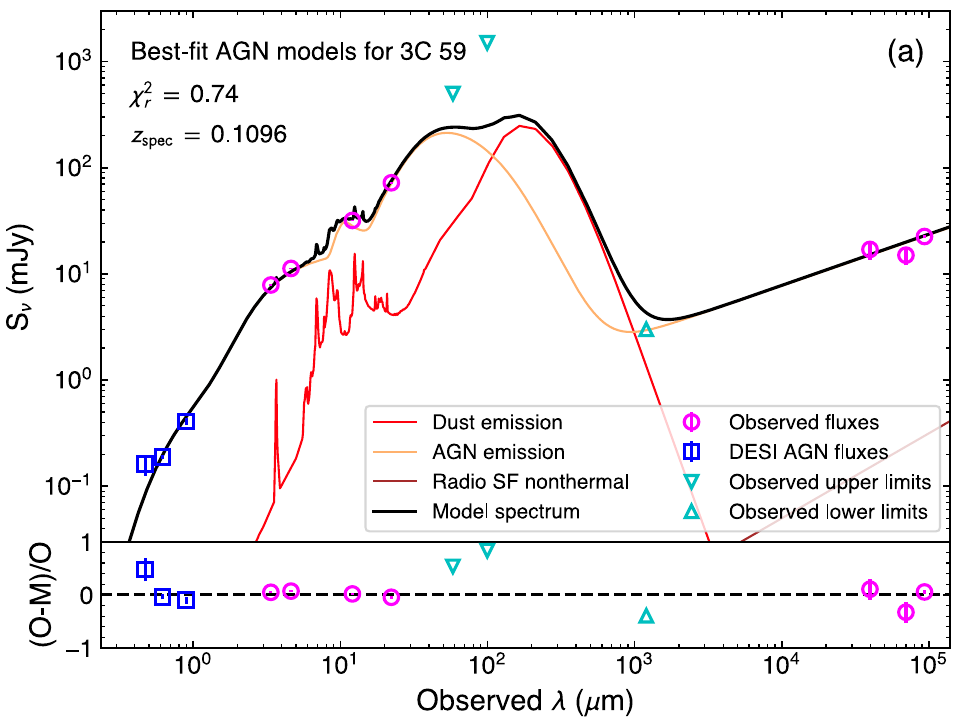}
\caption{Best-fit SED model for AGN and dust components of 3C 59 with \textsc{cigale}.
The observed fluxes and upper limits are shown in Table \ref{tab:3c59data}.
The DESI AGN fluxes are listed in Table \ref{tab:structurepara}.
\label{fig:3C59AGNSED}}
\end{figure}

% ########################      Table D2     ##############################
   \begin{table*}[h!]
   \caption{Model parameter settings in the entire SED fitting of 3C 59 \label{tab:3c59SEDpara}}
   \centering
   \scriptsize
   \setlength{\tabcolsep}{3pt}
   \begin{tabular}{clcccccc}
   \hline\hline
   Module & Parameter & Symbol & Values \\
   \hline
                                             & \multirow{2}{*}{Stellar $e$-folding time}  & \multirow{2}{*}{$\tau_{\rm{star}}$ ($10^6$ yr)}  &  1, 10, 50, 70, 100 to 950 (step 50), \\
Star formation history           &   &   &  1000 to 3000 (step 100), 3500 to 10000 (step 500) \\ 
$[\rm{SFR} \propto t \exp(-t/\tau)]$ & \multirow{2}{*}{Stellar age}     & \multirow{2}{*}{$t_{\rm{star}}$ ($10^6$ yr)}     & 50, 70, 100, 150, 170, 200, 220, 250, 300 to 1000 (step 50), \\
   &        &        &  1000, 1600, 2500, 3600, 5000, 7000, 10000  \\
  \hline
\textbf{Simple stellar population}      & Initial mass function      & --                                              & \cite{Chabrier2003} \\
 \citep{Bruzual2003}               & Metallicity                     & $Z$                                           & 0.0001, 0.0004, 0.004, 0.008, 0.02, 0.05 \\
  \hline
Dust attenuation  & \multirow{2}{*}{Color excess of the nebular lines} & \multirow{2}{*}{$E(B-V)$ (mag)}        & \multirow{2}{*}{0.005, 0.05, 0.1 to 0.5 (step 0.01), 0.55 to 1.5 (step 0.05)} \\
\citep{Calzetti2000} & & &   \\ 
\hline
Galactic dust emission & \multirow{2}{*}{Slope in $dM_{\rm{dust}} \propto U^{-\alpha}dU$} & \multirow{2}{*}{$\alpha$} & \multirow{2}{*}{3.5$^{*}$} \\
\citep{Dale2014} & & & \\
\hline
\multirow{2}{*}{AGN (UV-to-IR)} & AGN contribution to IR luminosity & $\rm{frac}_{\rm{AGN}}$ & 0.8$^{*}$ \\ 
          & Viewing angle & $\theta$ & $30^{\circ}$$^{*}$ \\
 \citep{Stalevski2012,Stalevski2016} & Polar-dust color excess & $E(B-V)_{\rm{PD}}$ (mag) & 0.8$^{*}$ \\
\hline
\multirow{3}{*}{Radio} & SF radio-IR correlation parameter & $q_{\rm{IR}}$ & 2.58$^{\dagger}$ \\
  & Radio-loudness parameter & $R_{\rm{AGN}}$ & 14$^{*}$  \\
  & Spectral index of AGN radio emission  & $\alpha_{\rm AGN}$ & 0.48$^*$ \\
\hline
   \end{tabular}
   \tablecomments{$^{\dagger}$ The $q_{\rm IR}$ is fixed to the default value of \textsc{cigale}. 
   $^{*}$ These parameters are fixed to the best-fit values obtained in the SED fitting with only AGN and dust models (see details in Section \ref{sec:3c59SED}).}
   \end{table*}

% ########################      Table D3     ##############################
   \begin{table*}[h!]
   \caption{Model parameter settings in the SED fitting for the five satellite galaxies \label{tab:compSEDpara}}
   \centering
   \scriptsize
   \begin{tabular}{clcccccc}
   \hline\hline
   Module & Parameter & Symbol & Values \\
   \hline
                                             & \multirow{2}{*}{Stellar $e$-folding time}  & \multirow{2}{*}{$\tau_{\rm{star}}$ ($10^6$ yr)}  &  1, 10, 50, 70, 100 to 950 (step 50), \\
Star formation history           &   &   &  1000 to 3000 (step 100), 3500 to 10000 (step 500) \\ 
$[\rm{SFR} \propto t \exp(-t/\tau)]$ & \multirow{2}{*}{Stellar age}     & \multirow{2}{*}{$t_{\rm{star}}$ ($10^6$ yr)}     & 50, 70, 100, 150, 170, 200, 220, 250, 300 to 1000 (step 50), \\
   &        &        &  1000, 1600, 2500, 3600, 5000, 7000, 10000  \\
  \hline
\textbf{Simple stellar population}      & Initial mass function      & --                                              & \cite{Chabrier2003} \\
 \citep{Bruzual2003}               & Metallicity                     & $Z$                                           & 0.0001, 0.0004, 0.004, 0.008, 0.02, 0.05 \\
  \hline
Dust attenuation  & \multirow{2}{*}{Color excess of the nebular lines} & \multirow{2}{*}{$E(B-V)$ (mag)}        & \multirow{2}{*}{0.005, 0.01, 0.025, 0.05, 0.075, 0.10 to 1.5 (step 0.05)} \\
\citep{Calzetti2000} & & &   \\ 
\hline
   \end{tabular}
   \end{table*}

% ######################################################################
% ########################        Appendix E      ##############################
% ######################################################################
\newpage
\section{AGN properties of 3C 59}
\label{sec:AGNproperties3C59}

\setcounter{table}{0}
\renewcommand{\thetable}{E\arabic{table}}

\setcounter{figure}{0}
\renewcommand{\thefigure}{E\arabic{figure}}

The physical properties of 3C 39 are summarize in Table \ref{tab:physicalpara}.
3C 59 has a spectral index of AGN radio emission with $\alpha_{\rm AGN} = -0.41\pm 0.12$,
which indicates that 3C 59 belongs to the flat spectrum radio quasars \citep[FSRQs; $\alpha_{\rm AGN} > -0.5$;][]{Mao2017}.
Total 8--1000 $\mu$m infrared (IR) luminosity is around $(1.22\pm 0.07)\times 10^{38}$ W, which is lower than the
maximum IR luminosity with $2.16\times 10^{38}$ W predicted by the 
3$\sigma$ upper limit of IRAS 100 $\mu$m flux upper limit \citep{Rothberg2013}.
The fraction of AGN IR luminosity to total IR luminosity ($F_{\rm AGN}$) is $80\% \pm 11\%$,
which indicates that the entire IR radiation of 3C 59 is dominated by the AGN emission.
The radio loudness estimated through the SED fitting in this work is $14 \pm 2$. 
The rest-frame 1.4 GHz radio luminosity ($L_{\rm 1.4 GHz, rest}$) is $8.9\times 10^{23}$ W Hz$^{-1}$ that is calculated by $L_{\rm 1.4 GHz, rest}=4\pi D_L^2/(1+z)^{1+\alpha_{\rm AGN}} \times (1.4/3)^{\alpha_{\rm AGN}}\times S_{\rm 3GHz, obs}$ where $S_{\rm 3GHz, obs}$ is the observed integrated flux densities at VLA 3 GHz \citep{Ceraj2018}.
\cite{Wang2024a} obtained the probability of radio AGNs hosted by SFGs or QGs
as a function of stellar mass, radio luminosity, and redshift.
At the stellar mass, radio luminosity, and redshift of 3C 59, the corresponding probability of hosting radio AGNs in QGs is almost 2.5 times higher than that in SFGs,
but 3C 59 is hosted by an SFG (see details in Section \ref{sec:rejuvenationSF3C59}). Therefore, it is a special case.

% ######################################################################
% ########################        Appendix F      ##############################
% ######################################################################
\newpage
\section{SED fitting with Prospector for 3C 59}
\label{sec:prospector3C59}
\setcounter{table}{0}
\renewcommand{\thetable}{F\arabic{table}}

\setcounter{figure}{0}
\renewcommand{\thefigure}{F\arabic{figure}}

The modules, parameters, and prior functions used in the SED fitting with \textsc{prospector} are shown in Table \ref{tab:3c59SEDprospector}.

% ########################      Table F1     ##############################
   \begin{table*}[h!]
   \caption{Parameters and priors for SED fitting with \textsc{prospector} for 3C 59 \label{tab:3c59SEDprospector}}
   \centering
   \scriptsize
   \setlength{\tabcolsep}{3pt}
   \begin{tabular}{clcccccc}
   \hline\hline
   Module & Parameter & Description & Prior function [prior range] \\
   \hline
   Star formation history & $\log (M_\star/M_\odot)$ & Total stellar mass formed & Uniform [7, 12] \\
                                      & $\log r_i$ & Ratio of the SFRs between adjacent bins & StudentT(0, 0.3, 2) \\
                                      & $\log (Z_\star/Z_\odot)$ & Stellar metallicity & Uniform [-2, 0.19] \\
   \hline
   Dust attenuation & dust1 & Attenuation by dust of stellar birth clouds & Uniform [0.1, 5] \\ 
                              & dust2 & Attenuation by dust of the diffuse interstellar medium & Uniform [0, 5] \\
                              & $\Gamma_{\rm dust}$ & Slope of \cite{Kriek2013} dust law & Uniform [-2, 0.5]\\
   \hline
   Nebular emission & $\log U_{\rm neb}$ & Ionization parameter of nebular emission & Uniform [-4, -1] \\
   \hline
   Dust emission & $U_{\rm min}$ & Minimum starlight intensity relative to the local interstellar radiation field & Uniform [0.1, 25] \\
                           & $Q_{\rm PAH}$ & Polycyclic aromatic hydrocarbon (PAH) mass fraction & Uniform [0.5, 7] \\
                           & $\gamma_{\rm dust}$ & Warm dust fraction & LogUniform [0.001, 0.15] \\
\hline
   \end{tabular}
   \end{table*}
 %-------------------------------------------------------------------------------------------------------------------------

\bibliography{ms}{}
\bibliographystyle{aasjournal}

\end{document}